\documentclass[twocolumn,showpacs,preprintnumbers,amsmath,amssymb]{revtex4}
\usepackage{graphicx}% Include figure files
\usepackage{dcolumn}% Align table columns on decimal point
\usepackage{bm}% bold math
\usepackage{latexsym}

\newcommand{\LSGE}       {linearized sine-Gordon equation}

\begin{document}

\title{Josephson Plasma Resonance in Bi$_2$Sr$_2$CaCu$_2$O$_{8+y}$ with Spatially Dependent Interlayer-Phase Coherence}
\author{N. Kameda,$^{1}$ M. Tokunaga ,$^{1}$ T. Tamegai,$^{1,2}$ M. Konczykowski${}^3$ and S. Okayasu${}^4$}
\address{${}^1$Department of Applied Physics, The University of Tokyo,
7-3-1 Hongo, Bunkyo-ku, Tokyo 113-8656, Japan}
\address{${}^2$CREST, Japan Science and Technology Corporation (JST)}
\address{${}^3$CNRS, URA 1380, Laboratoire des Solides Irradi\'{e}s,
         \'{E}cole Polytechnique, 91128 Palaiseau, France}
\address{${}^4$JAERI, 2-4, Shirakata Shirane, Tokai-mura, Naka-gun, Ibaraki, 319-1195, Japan}
         
\date{\today}

%\wideabs{
\begin{abstract}
We study the Josephson plasma resonance (JPR) in Bi$_2$Sr$_2$CaCu$_2$O$_{8+y}$ (BSCCO)
with spatially dependent interlayer-phase coherence (IPC).
The half-irradiated BSCCO (HI-BSCCO), in which columnar defects are introduced only in a half of the sample, shows several resonance peaks, which are not simple superposition of the peaks in irradiated- and pristine-parts.
JPR in HI-BSCCO changes its character from irradiated- to pristine-type at a crossover frequency ($\omega_{cr}$).
We demonstrate that the one-dimensional \LSGE, which takes into account the spatial dependence of IPC, can reproduce most of the experimental findings including the presence of $\omega_{cr}$.
\end{abstract}
%}wideabs 
%
\maketitle
%\section{INTRODUCTION}%%%%%%%%%%%%%%%%%%%%%%%%%%%%%%%%%%%%%%%%%%%%%%%%%%%%%%
Highly anisotropic high temperature superconductors like Bi$_2$Sr$_2$CaCu$_2$O$_{8+y}$ (BSCCO) can be considered as natural stacks of Josephson junctions.
In such systems, collective oscillation of superfluid density gives rise to the plasma oscillation along the stacking direction of $c$-axis \cite{Tachiki1994}.
A resonance occurs when the frequency of the incident electromagnetic field matches the plasma frequency.
This resonance is called Josephson plasma resonance (JPR) and its frequency is give by the following equation
\begin{eqnarray}
\omega_{p}^{2}=\omega_{p0}^{2} <\cos \phi_{n,n+1}>.
\end{eqnarray}
Here $\phi_{n,n+1}$ is the gauge-invariant phase difference between the neighboring layers which determines the interlayer-phase coherence (IPC), $\omega_{p0}$ ($=c/\lambda_{c}\sqrt{\epsilon_{0}}$) is the Josephson plasma frequency at zero field ($c$, $\lambda_{c}$, and $\epsilon_{0}$ are light velocity, $c$-axis penetration depth, and dielectric constant, respectively) \cite{Bula1995} and $<$ $>$ denotes thermal and disorder average \cite{daemen}.
JPR gives valuable information on the charge dynamics, superconducting symmetry, and vortex states \cite{Bula1995,tsui1994,matsuda1,shibauchi1999,sato1997,kosugi1997,hanaguri1997,kosherev1997} in anisotropic superconductors.

Most of JPR experiments, however, have been performed in macroscopically homogeneous systems.
What would happen when we have different coupling strengths in a system?
In this context, JPR in $T^{*}$-phase superconductors, which have two Josephson junctions with different coupling strengths in the unit cell, has been studied extensively both experimentally and theoretically \cite{marel,bulaevskii2002,shibata2,shibata1,kakeshita}.
Experimentally found two plasma modes were analyzed in the framework of Josephson junction stacks with alternating coupling strengths.
Another interesting question is what would happen if IPC has in-plane distribution on a macroscopic scale.
JPR frequency is greatly modified in the presence of vortices, since the configuration of pancake vortices determines $<\cos \phi_{n,n+1}>$ in Eq. (1).
When the pancake vortices are aligned along the $c$-axis, $<\cos \phi_{n,n+1}>$ is equal to unity and $\omega_{p}$ takes the maximum value.
A strong suppression of $<\cos \phi_{n,n+1}>$, and hence of $\omega_{p}$, occurs in magnetic fields due to the misalignment of pancake vortices.
Columnar defects (CDs), generated by heavy-ion irradiation, act as directional pinning centers and dramatically enhance $<\cos \phi_{n,n+1}>$ \cite{sato1997,kosugi1997,hanaguri1997}.
We can create an arbitrary distribution of IPC by controlling the pattern of heavy-ion irradiation.
In this paper, we report JPR in the simplest inhomogeneous system having two regions with and without CDs.
We find complicated field dependence of resonance peaks including their numbers.
We present a clear explanation of this non-trivial behavior of JPR through numerical calculations of one-dimensional sine-Gordon equation.

%\section{Experiments}%%%%%%%%%%%%%%%%%%%%%%%%%%%%%%%%%%%%%%%%%%%%%%%
  Single crystals of BSCCO are grown by using the traveling-solvent-floating-zone method \cite{ooi}.
The optimally-doped crystals ($T_{c}$= 90.0 K) are irradiated by 600 MeV iodine ions (A) (JAERI) and 6 GeV Pb ions (B, C and D) (GANIL) at the matching field of $B_{\Phi}$ = 20 kG.
The thickness of these crystals are small enough so that the heavy-ions can pass though.
The BSCCO crystal is half-irradiated (HI-BSCCO) by covering a half of the crystal with a gold foil, which is thick enough to stop the incident heavy-ions into the sample.
Dimensions of the samples are 670$\times$670$\times$20 $\mu$m$^{3}$ (A), 670$\times$670$\times$20 $\mu$m$^{3}$ (B), and 1480$\times$670$\times$20 $\mu$m$^{3}$ (C, D).
Samples C and D have the same dimensions but different distributions of CDs (see Fig.~3).
To study sample size dependence, sample B, C, and D are irradiated at the same dose.
JPR is measured by the cavity perturbation method at microwave frequencies of 24.2, 41.7, 46.3, 49.4, 52.0, 54.9, 56.6, and 61.2 GHz \cite{shibauchi1999}.
The crystal is set in a cylindrical copper cavity so that the microwave electric field is applied parallel to the $c$-axis.
The external magnetic field up to 90 kOe is also applied parallel to the $c$-axis by a superconducting magnet.
%\section{Results and discussions}%%%%%%%%%%%%%%%%%%%%%%%%%%%%%%%%%%%%%%%%%%%%%%%%%%%%%

Figure~1 (a) shows microwave power absorption as a function of field, $P_{abs}$($H$), for HI-BSCCO (A) at 50 K for various frequencies.
The dashed lines in Fig.~1(a) show JPR fields in irradiated and unirradiated samples with the same doping levels.
JPR in HI-BSCCO shows a complicated behavior, which is not a simple superposition of resonances from the constituent parts of the sample.
%The arrow ($\downarrow$) shows the presence of a tiny peak (see inset of Fig.~1(a)).
Change of frequency does not only shift the resonance field $H_{p}$, but also changes the number of resonance peaks.
At 24.2 GHz, only a single resonance peak appears close to the $H_{p}$ in the irradiated sample.
As frequency is increased from 24.2 GHz to 61.2 GHz, $H_{p}$ of the main peak shows characteristic anticyclotronic behavior \cite{matsuda1}.
Small shoulders observed at 41.7 and 46.3 GHz at the low field side of the main peak are most probably caused by original inhomogeneity of the crystal \cite{Yasugaki}.
Above a crossover frequency of $\omega_{cr}/2\pi$ $\sim$ 55 GHz, $H_{p}$ suddenly shifts to lower field.
%At the same time, new weak peaks ($\downarrow$) appear.
It would be worth mentioning that when we measure JPR for a pristine and an irradiated ($B_{\Phi}$ = 20 kG) samples placed side by side, the observed resonance peaks are simple superposition of the peaks in each individual sample (not shown).
This indicates that the electromagnetic continuity condition between pristine- and irradiated-part plays an important role to the behavior of JPR in HI-BSCCO.

Spatial and temporal dependence of the phase in a layered superconductor is described by a coupled sine-Gordon equation\cite{Bula92}.
According to Koshelev \cite{koshelevpra}, when the IPC along $c$-axis is uniform, spatial variation of the amplitude of plasma oscillation ($\theta(x)$) obeys the linearized sine-Gordon equation (LSGE)
\begin{equation}
\left(\frac{\omega^2+i\nu_{c}\omega}{\omega_{p0}^2}-C(x)\right)\theta(x)
+\lambda_{c}^{2}\nabla_{x}^{2}\theta(x)
=-\frac{i\omega D}{4\pi J_{0}},
\label{eq:Josephson}
\end{equation}
where $\omega$ and $C(x)$ represent incident microwave frequency and spatial dependence of IPC, respectively.
$\nu_{c}$, $D$, and $J_{0}$ ($=$ $\Phi_{0}\omega_{p0}^{2}\epsilon_{0}/8\pi^{2}s$) are damping frequency, microwave amplitude, and maximum Josephson current at zero field ($\Phi_{0}$ and $s$ are the flux quantum and the interlayer spacing).
In the present case, the spatial distribution of $C(x)$ is
\begin{equation}
    C(x)  = \begin{cases}
                           C_{1} & \text{$0 \leq x < L_{1}$} \\
                           C_{2} & \text{$L_{1} \leq x \leq L_{1}+L_{2}$ $(=L)$}
           \end{cases}.\nonumber
\end{equation}
Here, $L_{1}$ and $L_{2}$ are widths of pristine and irradiated parts (see inset of Fig.~1(b)) and $L$ is the total width of the sample.
Boundary conditions for Eq. (\ref{eq:Josephson}), $\theta'(0)=\theta'(L)=0$, are the requirement from no bias in-plane current at sample edges.
Microwave absorption power is originated from Joule-heating, and is given by
%\begin{equation}
$P(\omega) \propto \frac{\omega^2}{2}\int_{0}^{L} |\theta(x)|^2\mathrm{d}\it{x}$.
%\end{equation}
In the case of a homogeneous system where $C(x)$ in Eq. (\ref{eq:Josephson}) has a fixed value $C$ all over the sample, the resonance occurs at a single frequency of $\omega_{p}^{2}=C\omega_{p0}^{2}$.
Figure~1(b) shows the calculated absorption power as a function of external field.
In this calculation, field dependence of $C(x)$ for each part is assumed based on our previous results \cite{shibauchi1999}.
It should be noted, however, since the exact field dependence of IPC in the irradiated region is not known, we approximate it by three straight lines in the double logarithmic plot as shown by dotted lines in Fig.~2 \cite{kink20kOe}.
We also assume appropriate values of $\nu_{c}$, $\omega_{p0}$, $\epsilon_{0}$ and $\lambda_{c}$ as shown in Fig.~1(b).
A single resonance peak appears near the resonance field for the irradiated region below 45 GHz and the resonance peak shifts rapidly towards lower field between 50 GHz to 60 GHz, which we identify as $\omega_{cr}/2\pi$.
%We note that the small shoulders and the weak peaks at lower field side of the main peak in Fig.~1(a) are not shown in Fig.~2(a).
%They coused by an unexpected experimental disorders of $C(x)$.
%Ignoring these unintrinsic inhomogeneous effects, 
The LSGE reproduces all characteristic features of JPR in HI-BSCCO well.
Slight quantitative disagreements could be caused by the nonmonotonic field dependence of IPC in the irradiated part at low fields \cite{hanaguri1997,chikumoto}.
On the other hand, above $\omega_{cr}$, new weak peaks start to emerge at higher fields.
These peaks appear since Eq.~(2) without the damping term has the same form as the Schr\"{o}dinger equation.
The intensity of the new peaks becomes stronger at higher frequencies.
Since our measuring frequency is limited, we cannot observe the new peak in this sample.

Figure~2 shows calculated field dependence of $\omega_{p}$ in HI-BSCCO with the sample size of $L$ = 670 $\mu$m (sample A) and $L$ = 1480 $\mu$m (sample D).
We also add experimental data as solid circles in Fig.~2.
In this calculation, HI-BSCCO with $L$ = 670 $\mu$m shows three peaks in the frequency range, $0 \leq (\omega/\omega_{p0})^{2} \leq 1$.
In the case of isolated system, resonance peak in irradiated and the pristine parts are observed on $C_{1}$ and $C_{2}$ line, respectively.
Within our microwave frequency window, only the lowest frequency resonance is observed for $L$ = 670 $\mu$m sample.
Still, calculated results agree quantitatively with experimental data.
The calculated field dependence of the $\omega_{p}$ for HI-BSCCO with $L$ = 1480 $\mu$m is also shown with red lines.
Since doping levels in sample A and D are different, $C_{2}$ in sample A and D are shown as blue and red dotted lines, respectively.
To avoid complexity, we draw only two red lines as the lowest and the next lowest $\omega_{p}$ for $L$ = 1480 $\mu$m sample.
In addition, it is clear that $\omega_{cr}/2\pi$ is decreased to $\sim$ 30 GHz with increasing the sample size perpendicular to the boundary.

To confirm the size dependence of $\omega_{cr}$, we measure sample (B)-(D) with the same doping levels to each other.
Here C(and D), which have twice the size parallel (and perpendicular) to the irradiation boundary of sample (B), respectively.
Figure~3 shows $P_{abs}(H)$ in these samples for 41.7 GHz at 50 K.
The resonance in sample C occurs at almost the same field as in sample B.
This fact is valuable in further convincing ourselves that it is sufficient to consider only the one-dimensional equation as Eq.~(2). 
The resonance peak in sample D, which corresponds to the above calculation with $L$ = 1480 $\mu$m, shifts to lower field at 2.5 kOe and a new peak appear at 26 kOe.
In Fig.~2, we plot these peaks as closed squares.
The lower field peak and the new peak locate favorably on the lowest $\omega_{p}$ and the second lowest $\omega_{p}$ curve.
These results strongly suggest $\omega_{cr}/2\pi$ in sample D is reduced below 41.7 GHz region and support our claim that $\omega_{cr}$ tends to decrease with increasing the sample size perpendicular to the boundary.

Let us now discuss the behavior of JPR in HI-BSCCO in more detail.
For this purpose, we calculate the $\omega_{p}$ as a function of $C_{2}$ with fixed $C_{1}$.
Figure~4(a) shows the $\omega_{p}$ as a function of $C_{2}^{-1}$ (0.001 $\leq$ $C_{2}$ $\leq$ 1) with fixed $C_{1}$ = 0.001.
Since IPC changes more significantly in the irradiated part in our experiments, this figure approximates the actual field dependence of $\omega_{p}$.
Dotted lines in Fig. 4(a) represent the conditions of $\omega_{p}^{2}/\omega_{p0}^{2} = C_{1}$ and $C_{2}$.
At the intersection of the two lines (1), namely $C_{1}$ = $C_{2}$ = 0.001, HI-BSCCO becomes uniform and shows only one resonance at $\omega_{p}^{2} = C_{1}\omega_{p0}^{2} = 0.001 \omega_{p0}^2$.
The increase of $C_{2}$ causes an increase of $\omega_{p}$ ($\bullet$), and in addition, brings two new resonance peaks at higher frequencies ($\blacksquare$) and ($\blacktriangle$).

In order to get further insight into the nature of the resonances, we calculate spatial variation of the real part of $\theta(x)$ (Re[$\theta(x)$]).
The Eq. (2) is now analogous to the Schr\"{o}dinger equation for a particle in spatially dependent potential well with modified boundary conditions.
The Re[$\theta(x)$] shows a snapshot of the resonance state in the sample.
Figure~4(b) shows Re[$\theta(x)$] normalized by its maximum value in HI-BSCCO at several values of $C_{2}^{-1}$ and $\omega_{p}$ ((1) to (6) in Fig.~4(a)).
Re[$\theta(x)$] is constant for homogeneous case of (1) as expected, whereas it shows spatial variation for $C_{1} \neq C_{2}$ as in (2)-(6).
When $C_{2}$ is slightly larger than $C_{1}$ as in (3) and (6), Re[$\theta(x)$] changes sinusoidally and is antisymmetric (or symmetric) with respect to the center of the sample.
As $C_{2}$ increases further, sinusoidal variation of Re[$\theta(x)$] is modified.
When $\omega_{p}^{2}/\omega_{p0}^{2}$ is equal to $C_{2}$ as in (4), Re[$\theta(x)$] changes sinusoidally only in the pristine part and is nearly constant in the irradiated part, since the wave length of Re[$\theta(x)$] in the irradiated part diverges to infinity.
In the case of $C_{1} < \omega_{p}^{2}/\omega_{p0}^{2} < C_{2}$, the spatial variation of Re[$\theta(x)$] in the irradiated part is damped as shown in (2) and (5).
The finite amplitude of Re[$\theta(x)$] in the irradiated part is the result of the continuity condition at the boundary.

To clarify the origin of additional resonance, we calculate the $\omega_{p}$ ignoring the damping term in Eq.~(2), which has little influence on the determination of $\omega_{p}$.
The resulting equation has the same form as the Schr\"{o}dinger equation for a particle in a box.
In an uniform system ($C_{1} = C_{2}$), spatial derivative of $\theta$ is neglected, and we have only one resonance at $\omega_{p}^{2}/\omega_{p0}^{2} = C_{1} = C_{2}$.
In case of $\omega_{p}^{2}/\omega_{p0}^{2}>C_{1} \neq C_{2}$, $\theta(x)$ is represented by the sum of transverse- and longitudinal-components as 
\begin{equation}
\theta_{j}(x) = A_{j}e^{ik_{j}x}+B_{j}e^{-ik_{j}x}-\frac{i\omega D}{4\pi J_{0}}\left(\frac{\omega^2}{\omega_{p0}^2}-C_{j}\right)^{-1}
\nonumber
\end{equation}
with ($j$=1, 2).
Here $k_{j}$ $(=(1/\lambda_{c})\sqrt{\omega^{2}/\omega_{p0}^{2}-C_{j}})$ is the wave number.
In the case of $C_{1} \sim C_{2}$, $\theta(x)$ is approximately expressed by a sinusoidal wave as in (3) and (6), since $k_{1}$ and $k_{2}$ are nealy equal.
Then $k_{1}$ and $k_{2}$ must satisfy the condition $\sin(k_{1}L )\sim \sin(k_{2}L)$ = 0 by the boundary condition of Eq.~(2), from which we can evaluate $\omega_{p}$ as $\omega_{n}$, where
\begin{equation}
\frac{\omega_{n}^{2}(C_{1}, L)}{\omega_{p0}^{2}} = C_{1}+(\frac{\lambda_{c}}{L}n\pi)^2
\end{equation}
with $n = 0, 1, 2, \cdot \cdot \cdot$.
In the case of $C_{2} > \omega_{p}^{2}/\omega_{p0}^{2} > C_{1}$, $\omega_{p}$ can be obtained from the intersection of two lines
$k_{1}\tan(k_{1}L/2) =\kappa_{2} \tanh(\kappa_{2}L/2)$
and $\kappa_{2}^{2}+k_{1}^{2}=C_{2}-C_{1}$.
Here $\kappa_{2}=(1/\lambda_{c})\sqrt{C_{2}-\omega^{2}/\omega_{p0}^{2}}$ is the attenuation coefficient in the irradiated part.
In the limit of $C_{2} \gg C_{1}$,
the two lines cross to each other at $k_{1}L/2 = \pi/2+n\pi$, which translates into $k_{1} = \pi(2n+1)/L$.
This means that $\omega_{p}$ started from $\omega_{n}$ ($n = 0, 1, 2, \cdot \cdot \cdot$) approaches to $\omega_{2n+1}(C_{1}, L)$.
Dotted lines in Fig.~4(a) indicate $\omega_{p} = \omega_{n}(C_{1}, L)$ for $n=1, 2, 3$ with $L=$ 670 $\mu$m and $C_{1}=$ 0.001.
Actually, in the limit of $C_{2} \rightarrow C_{1}$, each resonance frequencies ($\bullet$, $\blacksquare$, and $\blacktriangle$) approaches to $\omega_{n}$ with $n = 0, 1,$ and 2.
By increasing $C_{2}$, $\omega_{p}$ started from $\omega_{n}$ shift towards higher frequency and it approaches to $\omega_{2n+1}$ in the limit of large $C_{2}$.
It should be noted here that in the real system, the maximum of $C_{2}$ is limited by unity.
%and hence $\omega_{p}$ near saturate for $n = 1$ and 2.
Therefore, JPR frequencies in HI-BSCCO can be classified by characteristic frequencies $\omega_{n}(C_{1}, L) (n=0, 1, 2, \cdot \cdot \cdot)$.

Finally, we discuss the origin of $\omega_{cr}$.
The observed plasma resonance in HI-BSCCO with $L$ = 670 $\mu$m corresponds to the lowest $\omega_{p}$ ($\bullet$) in Fig.~4(a), since we measure at relatively low frequency.
In the calculation with fixed $C_{1}$ shown in Fig.~4(a), the lowest $\omega_{p}$ approaches to $\omega_{1}(C_{1}, L)$ in the limit of $C_{2} \gg C_{1}$ at lower fields.
However in reality, the limiting value $\omega_{1}$ itself is a function $C_{1}$($\propto 1/H$).
Considering the field dependence of $C_{1}$, we calculate $\omega_{1}^{2}(C_{1}(H), L)/\omega_{p0}^{2}$ for $L$ = 670 and 1480 $\mu$m, which are shown in Fig.~2 by dash-dotted lines.
Now it is clear that the crossover frequency $\omega_{cr}$ appears since $\omega_{p}^{2}$ follows $\omega_{1}^{2}(C_{1}(H), L)$ at lower fields when $C_{2}$ is almost saturated.
On increasing field, $\omega_{p}(H)$ starts to follow the lowest line in Fig.~4 since the change in $C_{1}$ is much less than that in $C_{2}$.

%\section{Summary}%%%%%%%%%%%%%%%%%%%%%%%%%%%%%%%%%%%%%%%%%%%%%%%%%%%%%%%%

In summary, we study the Josephson plasma resonance in half-irradiated Bi$_2$Sr$_2$CaCu$_2$O$_{8+y}$(HI-BSCCO) with inhomogeneous phase coherence.
All experimental results, including multiple resonance and the crossover frequency, are well reproduced by calculations based on the \LSGE.
Resonance peaks in HI-BSCCO are classified in terms of characteristic frequencies, which belong to different modes with specific wavelength relative to the sample size.
In such layered superconductors with spatially dependent interlayer-phase coherence, microwave electric field parallel to the $c$-axis can partly excite the transverse mode, which can only be excited by microwave magnetic field in homogeneous systems.
%\section{ACKNOWLEDGEMENTS}%%%%%%%%%%%%%%%%%%%%%%%%%%%%%%%%

We are grateful to A. E. Koshelev, T. Koyama and M. Machida for stimulating discussion.
This work is supported by CREST and Grant-in-Aid for Scientific Research from Ministry of Education, Culture, Sports, Science, and Technology, Japan.

\newpage
.
\newpage
% %%%%%%% Fig. 1 %%%%%%%%%%%%%%%%%%%%%%%%%%%%%%%%%%
\begin{figure}
\begin{center}\includegraphics[width=130mm]{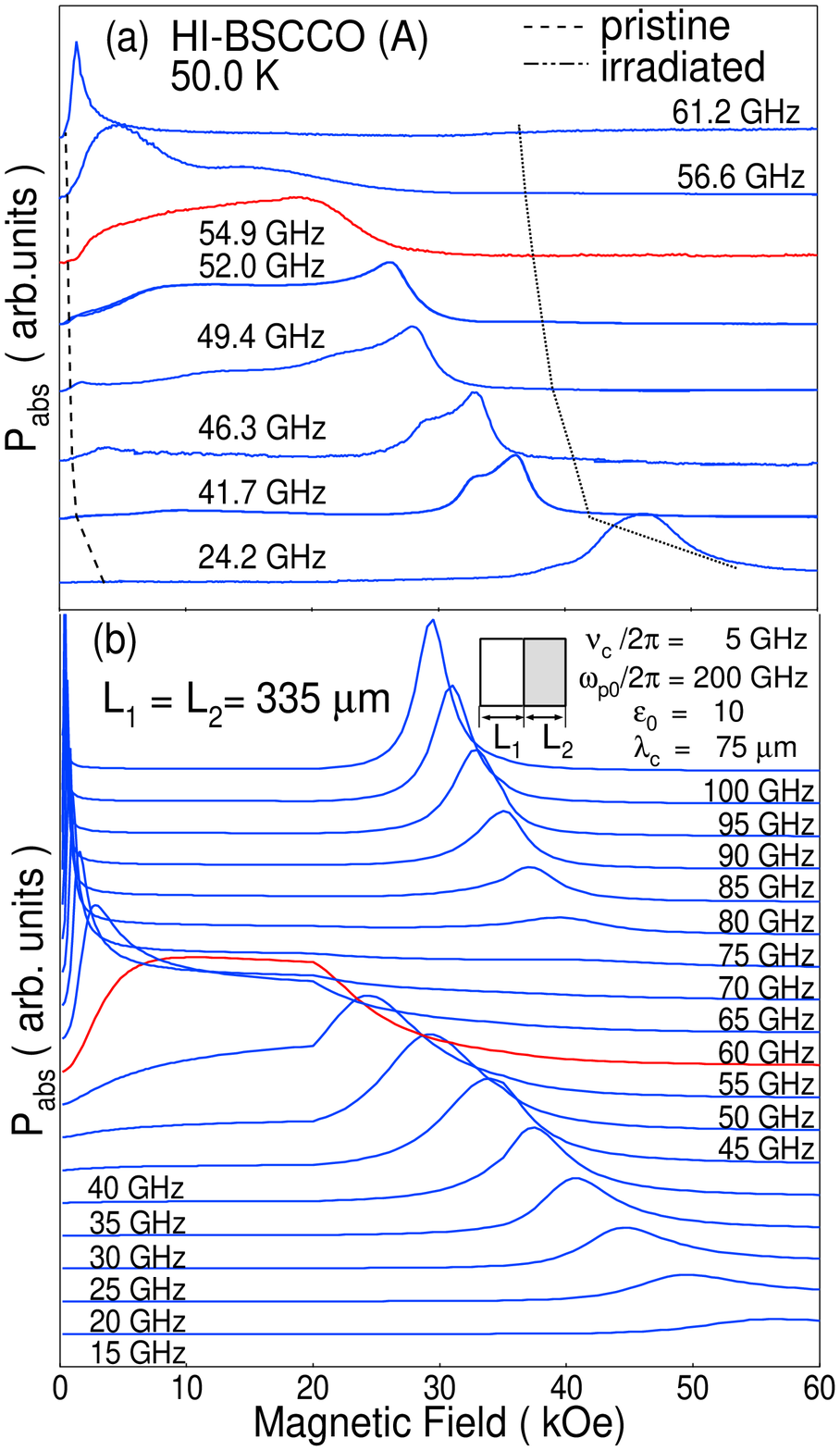}
\end{center}
\caption{
(a) $P_{abs}(H)$ for HI-BSCCO at 24.2, 41.7, 46.3, 49.4, 52.0, 54.9, 56.6, and 61.2 GHz  at 50 K.
Dashed and dotted lines indicate the resonance fields for pristine and irradiated part of the sample, respectively.
(b) $P_{abs}(H)$ calculated based on the linearized sine-Gordon equation (Eq.~(2)) for position dependent IPC.
}
\label{fig1}
\end{figure}
% %%%%%%%%%%%%%%%%%%%%%%%%%%%%%%%%%%%%%%%%%%%%%%%%%%

% %%%%%%% Fig. 2 %%%%%%%%%%%%%%%%%%%%%%%%%%%%%%%%%%
\begin{figure}
\begin{center}\includegraphics[width=130mm]{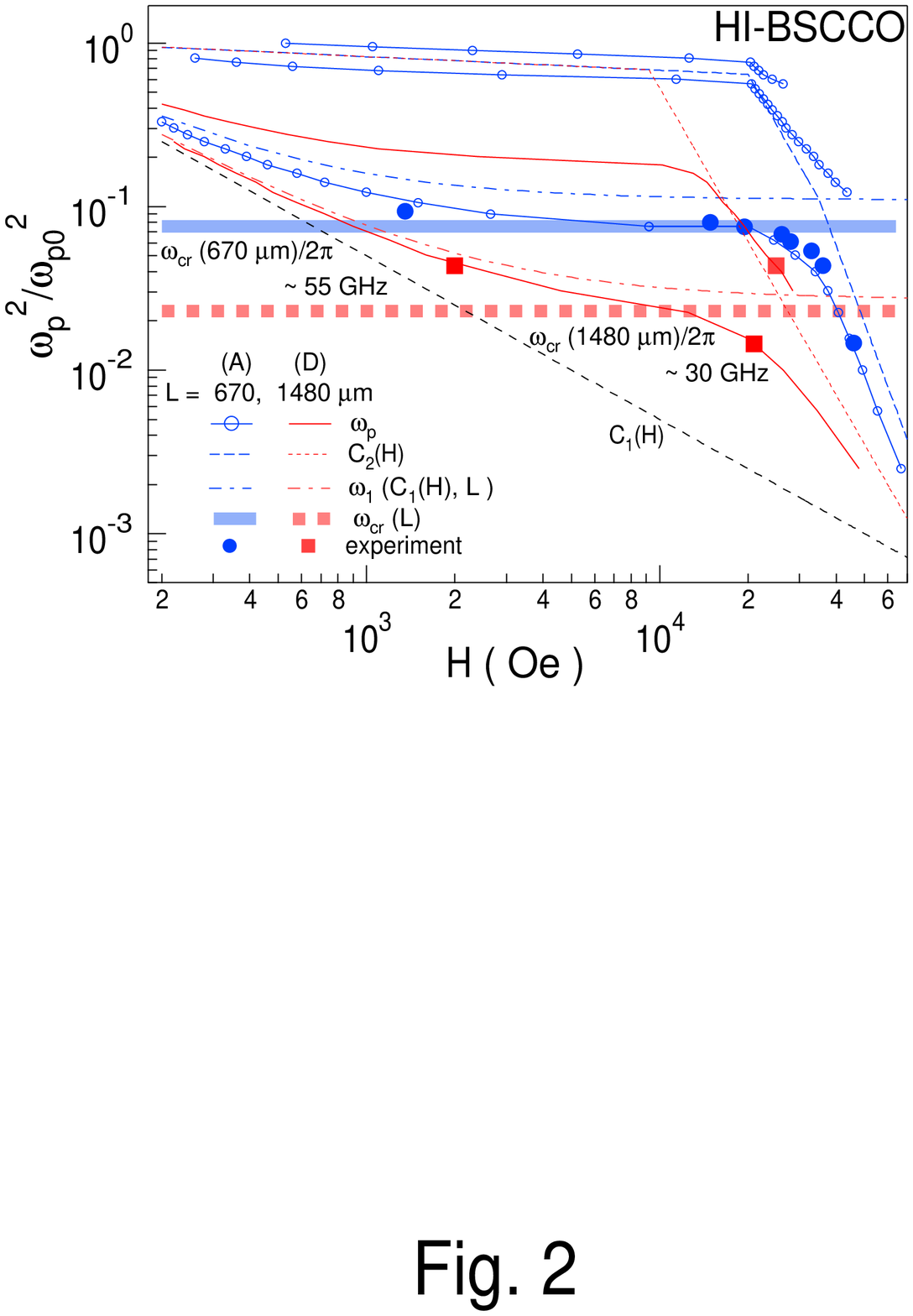}
\end{center}
\caption{
(Color online) Field dependence of $\omega_{p}$ for HI-BSCCO with $L$= 670 ($\circ$) and 1480 $\mu$m (red lines) calculated by \LSGE.
Solid circles ($\bullet$) and Solid squares ($\blacksquare$) show observed resonance peaks at 50 K.
The dashed line and dotted lines show $C_{1}(H)$ and $C_{2}(H)$ which are used in this calculation, respectively. 
For explanation of $\omega_{cr}$ and $\omega_{1}(C_{1}, L)$, see text and Eq.~(3).
}
 \label{fig2}
\end{figure}
% %%%%%%%%%%%%%%%%%%%%%%%%%%%%%%%%%%%%%%%%%%%%%%%%%%
% %%%%%%% Fig. 3 %%%%%%%%%%%%%%%%%%%%%%%%%%%%%%%%%%
\begin{figure}
\begin{center}\includegraphics[width=130mm]{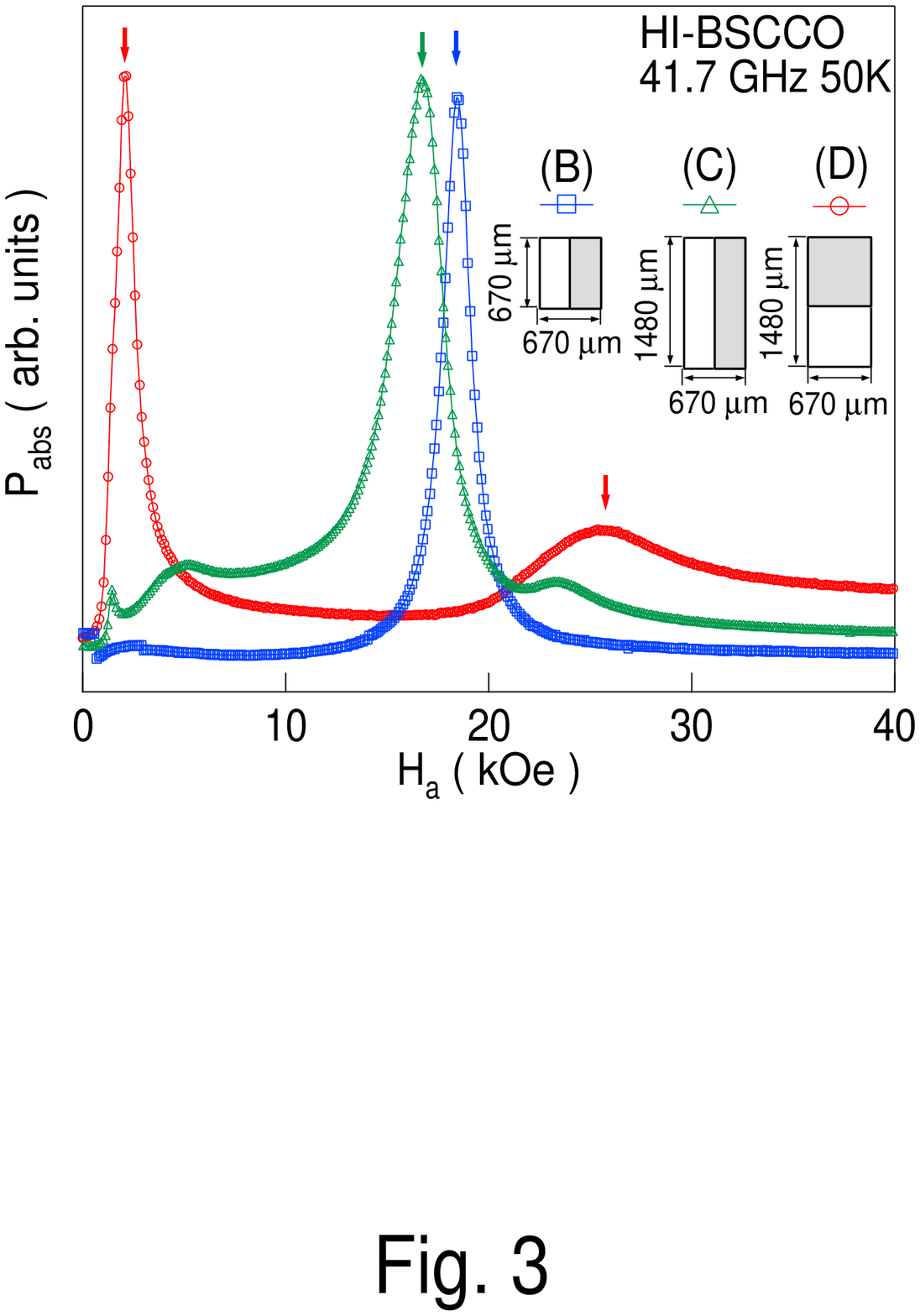}
\end{center}
%\resizebox{70mm}{!}{\includegraphics[width=140mm]{Graph/Layout3.eps}}
\caption{
$P_{abs}(H)$ in three HI-BSCCO samples with different dimensions and irradiation patterns with $B_{\Phi}$ = 20 kG for 41.7 GHz at 50 K.
Resonance peaks are marked by arrows.
The irradiation pattern for each sample is schematically shown.
The difference between samples C and D is the direction of the irradiation boundary.
}
\label{fig3}
\end{figure}
% %%%%%%%%%%%%%%%%%%%%%%%%%%%%%%%%%%%%%%%%%%%%%%%%
% %%%%%%% Fig. 4 %%%%%%%%%%%%%%%%%%%%%%%%%%%%%%%%%%
\begin{figure}
\begin{center}\includegraphics[width=130mm]{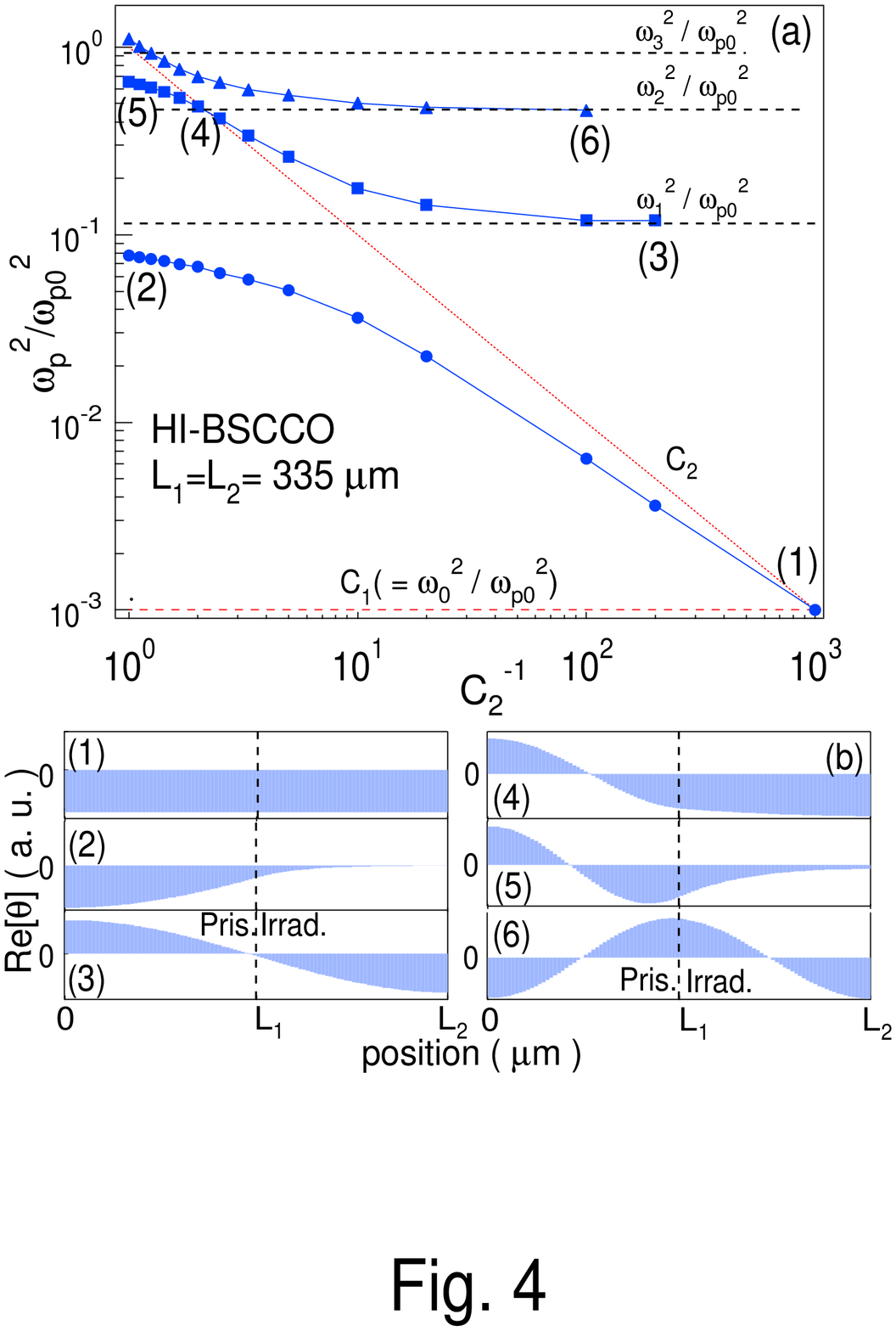}
\end{center}
\caption{
(a) $C_{2}$ dependence of the peak frequency ($\omega_{p}$) (solid symbols).
$C_{1}$ is fixed at 0.001.
Dotted lines represent $C_{1}$ and $C_{2}$.
Dashed lines represent $\omega_{n}^{2}/\omega_{p0}^{2}$ $(n=0,1,2,$ and $3)$ with $C_{1} =$ 0.001 and $L_{1}+L_{2} =$ 670 $\mu$m.
Here, $\omega_{n}=\omega_{n}(C_{1},L)$.
(b) 
The distribution of Re[$\theta(x)$] in sample at various points in (a).
}
\label{fig4}
\end{figure}
% %%%%%%%%%%%%%%%%%%%%%%%%%%%%%%%%%%%%%%%%%%%%%%%%%%

%%%%%%%%%%%%%% bibliography %%%%%%%%%%%%%%%%%%%%%%%%%%%%%%%%%%%%%%
%
% now the references. delete or change fake bibitem. delete next three
%   lines and directly read in your .bbl file if you use bibtex.
% \begin{figure}
% \caption{}
% \label{}
% \end{figure}
% \begin{table}
% \caption{}
% \label{}
% \end{tabular}
% \end{table}

\begin{references}
\bibitem{kink20kOe}

The kink at 20 kOe for all frequency is an artifact caused by the assumed abrupt change in the slope of $C_{2}(H)$ at this field (see Fig.~2).
\end{references}
\end{document}